\documentclass[12pt]{iopart}%
\usepackage{graphicx}

\begin{document}
\title[Simultaneous measurement of coordinate and momentum...]{Simultaneous measurement of coordinate and momentum on a von Neumann
lattice}
\author{A Mann, M Revzen and J Zak}
\address{Department of Physics,
Technion-Israel Institute of Technology, Haifa 32000, Israel}
\ead{\mailto{ady@physics.technion.ac.il},\mailto{revzen@physics.technion.ac.il},\mailto{zak@physics.technion.ac.il}}

\begin{abstract}
It is shown that on a finite phase plane the $kq$-coordinates and the sites of
a von Neumann lattice are conjugate to one another. This elementary result
holds when the number $M$ defining the size of the phase plane can be
expressed as a product, $M=M_{1}M_{2}$, with $M_{1}$ and $M_{2}$ being
relatively prime. As a consequence of this result a hitherto unknown wave
function is defined giving the probability of simultaneously measuring the
momentum and coordinate on the von Neumann lattice.
\end{abstract}


Following the pioneering work of Schwinger [1] in the finite phase plane,
there has been much interest and activity in a great variety of physical
problems, including quantum measurements [2], quantum computing [3], quantum
maps [4], Landau levels in a magnetic field and von Neumann lattices [5],
quantum teleportation [6], and others (see review article by A. Vourdas [7]).
In general, when limiting the motion to a finite phase plane, one might obtain
some surprising results that do not exist in the infinite phase plane. An
example of such a result are the mutually unbiased bases [8] which are
characteristic for a finite phase plane. Another example is the Wigner
function which requires different definitions for even and odd dimensionality
$M$ of the phase plane [9]. Yet another example is the $kq$-representation in
finite phase plane [5,10]. In this example, when $M=M_{1}M_{2}$, with $M_{1}$
and $M_{2}$ being relatively prime, one can construct a $KQ$-representation,
which is conjugate to the $kq$-representation. This is not possible in an
infinite phase plane.

In this Letter we show that in a finite phase plane with lengths $Mc$ in the
$x-$ direction and $\hbar\frac{2\pi}{c}$ in the $p$-direction ($c$ is a
constant), the $kq$-coordinates [11] and the sites of the corresponding von
Neumann lattice are mutually conjugate. This result holds when $M=M_{1}M_{2}$,
with $M_{1}$ and $M_{2}$ being relatively prime. The $kq$-coordinates are
defined by the eigenvalues of the commuting operators $T(a)=\exp\left(
\frac{i}{\hbar}pa\right) $ and $\tau\left( \frac{2\pi}{a}\right) =\exp\left(
ix\frac{2\pi}{a}\right) $, where $a=M_{1}c$. The unit cell of the
corresponding von Neumann lattice has the dimensions $a$ and $\hbar\frac{2\pi
}{a}$in the $x$ and $p$-directions, respectively. We call it the $a$-von
Neumann lattice. An additional set of $KQ$-coordinates is built which are
eigenvalues of the commuting operators $T(b)=\exp\left( \frac{i}{\hbar
}pb\right) $ and $\tau\left( \frac{2\pi}{b}\right) =\exp\left( ix\frac{2\pi
}{b}\right) $, where $b=M_{2}c$ [10]. Similarly, the $KQ$-coordinates are
conjugate to the sites of the $b$-von Neumann lattice, having as the unit cell
$b$ and $\hbar\frac{2\pi}{b}$, respectively. Then a very surprising and
unexpected result is proven, namely, that the eigenvalues of the operators
$T(a)$ and $\tau(\frac{2\pi}{a})$ are also the sites $u^{\prime}b$ and
$v^{\prime}\hbar\frac{2\pi}{b}$ of the $b$-von Neumann lattice, where
$u^{\prime}$ and $v^{\prime}$ are integers modulo $M_{1}$ and $M_{2}$,
respectively. A similar proof is given for the $KQ$-coordinates and the sites
of the $a$-von Neumann lattice. Having proven this we show how to define a new
wave function which gives the probability of a simultaneous measurement of
the momentum and coordinate on the von Neumann lattice.

As is known [5,10] a finite phase plane can be achieved by using boundary
conditions on the wave function $\psi(x)$ and its Fourier transform $F(p)$
\begin{equation}
\psi(x+Mc)=\psi(x) \ ; \ \ F\left( p + \hbar\frac{2\pi}{c}\right) =F(p),
\end{equation}
where $M$ is an integer and $c$ is a constant which determines the
discreteness of $x$ and $p$ in the phase plane. Following the boundary
conditions in Eq.\ (1), the coordinate $x$ and the momentum $p$ turn out to be
discrete and assume the values
\begin{equation}
x=sc, \ s=0,1, \dots, M-1 \ ; \ \ p = \hbar\frac{2\pi}{Mc} t, \ t= 0,1,\dots,
M-1 \ .
\end{equation}
In the finite phase plane the operators $x$ and $p$ are then replaced by the
following exponential operators
\begin{equation}
\tau\left( \frac{2\pi}{Mc}\right)  = \exp\left( ix\frac{2\pi}{Mc}\right)  \ ,
\ \ T(c)=\exp\left( \frac{i}{\hbar}pc\right)
\end{equation}
Let us assume that
\begin{equation}
M=M_{1}M_{2} \ .
\end{equation}
One then can define a $kq$-representation, based on the constant $a=M_{1}c$,
with the commuting operators [11]
\begin{equation}
T(a)=\exp\left( \frac{i}{\hbar}pa\right)  \ \mathrm{and} \ \tau\left(
\frac{2\pi}{a}\right) =\exp\left( ix\frac{2\pi}{a}\right)
\end{equation}
The eigenvalues of these operators are the $kq$-coordinates in the finite
phase plane $k_{n}$ and $q_{m}$
\begin{equation}
T(a):\exp(ik_{n}a) \ \mathrm{and} \ \tau\left( \frac{2\pi}{a}\right)
:\exp\left( iq_{m}\frac{2\pi}{a}\right)  \ ,
\end{equation}
where
\begin{equation}
k_{n}=\frac{2\pi}{Mc}n \ , \ \ n=0,1,\dots,M_{2}-1 \ ; \\ q_{m} =
mc, m=0,1, \dots M_{1}-1 \ \nonumber .
\end{equation}
For the case where $M=15$, $M_{1}=3$ and $M_{2}=5$, these eigenvalues are
shown in Fig. 1 by open circles. According to the definition of the
eigenvalues in Eqs.\ (6) and (7), $k_{n}$ is given modulo $\frac{2\pi}{a}$,
and $q_{m}$ is given modulo $a$. This means that any of the unit cells of
the
von Neumann lattice in Fig.\ 1 could be used for plotting the eigenvalues in
Eqs.\ (6) and (7). We would like to draw to the attention of the reader that
the area of the unit cell of the von Neumann lattice is $h$, the Planck
constant, which makes the von Neumann lattice very attractive.

Similarly, we can define another von Neumann lattice with the constant
$b=M_{2}c$, and respectively another $KQ$-representation (here we use capital
$K$ and $Q$) based on the commuting operators [10]
\begin{equation}
T(b)=\exp\left( \frac{i}{\hbar}pb\right)  \ \mathrm{and} \ \tau\left(
\frac{2\pi}{b}\right)  = \exp\left( i x \frac{2\pi}{b}\right)
\end{equation}
The eigenvalues of these commuting operators will be labelled by
$K_{n^{\prime}}$ and $Q_{m^{\prime}}$ which will assume the values [cf.
Eq.\ (7)]
\begin{equation}
K_{n^{\prime}}=\frac{2\pi}{Mc}n^{\prime}, \ n^{\prime}=0,1,\dots,M_{1}-1,
\\ Q_{m^{\prime}}=m^{\prime}c,m^{\prime}=0,1,\dots,M_{2}-1 \nonumber
\end{equation}
It is clear that for $M=15$ one can draw a figure similar to Fig.\ 1, but with
$a=3c$ replaced by $b=5c$. Correspondingly, we will have 3 unit cells in the
$x$-direction and 5 unit cells in the $p$-direction, see Fig.\ 2. Up to now
there was no restriction on the factors $M_{1}$ and $M_{2}$ in the partition
of $M$ in Eq.\ (4). However, as was shown in Ref.\ [10], a very basic new
result is obtained when $M_{1}$ and $M_{2}$ are relatively prime. Then the
operators in Eqs.\ (5) and (8) are mutually conjugate. In operator language
this conjugacy is expressed in the following way [10].
\begin{eqnarray}
T(a)\tau\left( \frac{2\pi}{b}\right) =\tau\left(
\frac{2\pi}{b}\right)
T(a)\exp\left( 2\pi i\frac{M_{1}}{M_{2}}\right) \\
T(b)\tau\left( \frac{2\pi}{a}\right) =\tau\left( \frac{2\pi}%
{a}\right) T(b)\exp\left( 2\pi i\frac{M_{2}}{M_{1}}\right)\nonumber
\end{eqnarray}
On the other hand, in the language of the eigenstates of the operators in
Eqs.\ (5) and (8) this conjugacy assumes the following form [10]
\begin{equation}
|\langle k,q | K, Q\rangle|^{2}=\frac{1}{M} \ ,
\end{equation}
where we have omitted the subscripts on $k,q,K$ and $Q$ [see Eqs.\ (7) and
(9)]. In Eq.\ (11) $|k,q\rangle$ are the eigenvectors of the commuting
operators in Eq.\ (5), while $|K,Q\rangle$ are the eigenvectors of the
commuting operators in Eq.\ (8).

We are now ready to prove that the $kq$-coordinates and the sites of the
$a$-von Neumann lattice are mutually conjugate. Here the $a$-von Neumann
lattice is the lattice that corresponds to the $kq$-representation [see
Eq.\ (5)] defined by means of the constant $a$. In the example for $M=15$,
this is given in Fig.\ 1. Let us denote the sites of the $a$-von Neumann
lattice by $\alpha_{st}$
\begin{equation}
\alpha_{st}=sa+t\hbar\frac{2\pi}{a} \ ,
\end{equation}
where $s$ runs from 0 till $M_{2}-1$, and $t$ from 0 till $M_{1}-1$. In our
example of $M=15$, $M_{1}=3$ and $M_{2}=5$, $s$ runs from 0 to 4 and $t$ runs
from 0 to 2 (see Fig.\ 1). At this juncture we have that the couples $kq$ and
$KQ$ are mutually conjugate coordinates. In order to prove that $kq$ and
$\alpha_{st}$ are mutually conjugate, we have to show that there is a
one-to-one correspondence between the coordinates $KQ$ and $\alpha_{st}$. Let
us consider the $K_{n^{\prime}}$ and $Q_{m^{\prime}}$ coordinates in Eq.\ (9)
and write the following equations
\begin{equation}
n^{\prime}\frac{2\pi}{Mc}+v^{\prime}\frac{2\pi}{b}=t\frac{2\pi}{a},
\ \ \ \ m^{\prime}c+u^{\prime}b=sa,
\end{equation}

\begin{figure}[th]
\begin{center}
\includegraphics[width=0.45\textwidth]{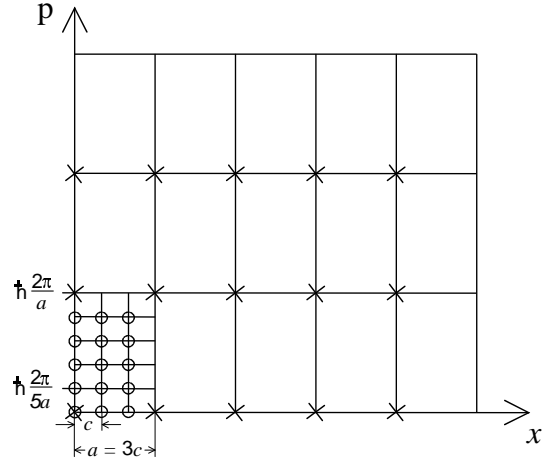}
\caption{von Neumann lattice (denoted by $\times$) and eigenvalues
$k_{n}$ and $q_{m}$ in Eq. (7) (denoted by open circles) of the
operators in Eq. (5), for $M=15, \ a=3c$.} \label{fig:1}
\end{center}
\end{figure}

\begin{figure}[th]
\begin{center}
\includegraphics[width=0.45\textwidth]{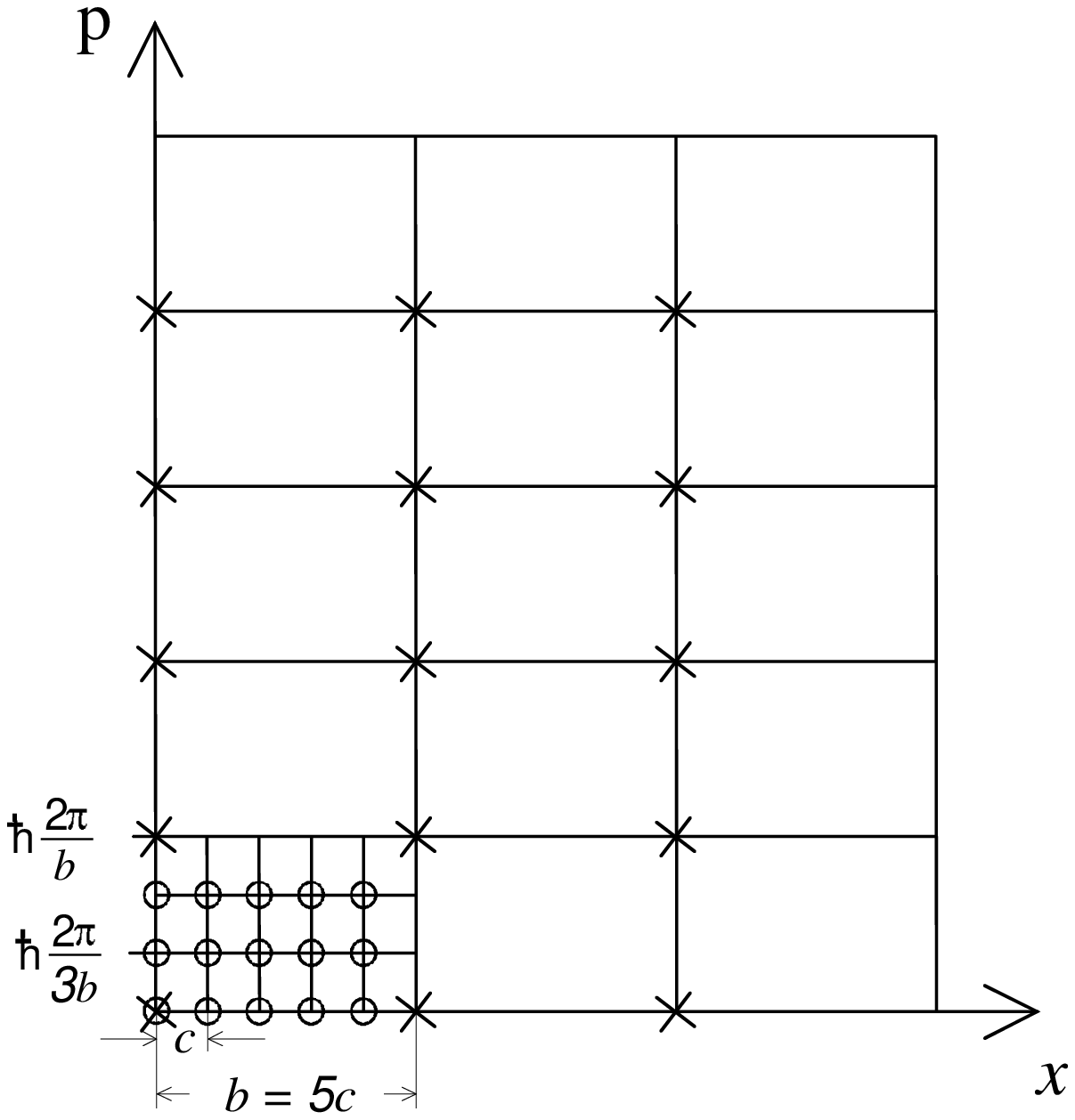}
\caption{von Neumann lattice (denoted by $\times$) and eigenvalues
$K_{n^{\prime}}$ and $Q_{m^{\prime}}$ in Eq. (9) (denoted by open
circules) of the operators in Eq. (8), for $M=15$, $b=5c$.}
\label{fig:2}
\end{center}
\end{figure}
where $s,t,u^{\prime}$ and $v^{\prime}$ are integers. In what
follows we shall use primed letters to relate to the $b$-von Neumann lattice
(unprimed are for the $a$-lattice). The meaning of Eqs.\ (13) is very simple.
On the left hand sides we have a $K_{n^{\prime}},Q_{m^{\prime}}$-point in the
unit cell of the $b$-von Neumann lattice modulo the $(v^{\prime},u^{\prime}%
)$-lattice site [see Eq.\ (9) and Fig.\ 2]. On the right hand side we have a
$(t,s)$-site of the $a$-von Neumann lattice. We can now convince ourselves
that the $K_{n^{\prime}},Q_{m^{\prime}}$-point determines uniquely the
$(t,s)$-site. For this let us rewrite the first equation of Eqs.\ (13) by
dividing both sides of it by $\frac{2\pi}{Mc}$. We get the following
Diophantine equation
\begin{equation}
n^{\prime}+M_{1}v^{\prime}=M_{2}t
\end{equation}
The particular feature of this equation is that for any $n^{\prime}%
=0,1,\dots,M_{1}-1$ there is a single well determined pair $(v^{\prime},t)$
solving it [10]. This means that there is a well defined integer $t$ (mod
$M_{1}$) for any $n^{\prime}$. Similarly, the second equation in Eqs.~(13),
after dividing by $c$, turns into
\begin{equation}
m^{\prime}+ M_{2}u^{\prime}=M_{1}s
\end{equation}
Again, this equation gives a well defined pair $(u^{\prime},s)$ for every
$m^{\prime}$. From Eqs.\ (13)-(15) it follows that the $K_{n^{\prime}%
},Q_{m^{\prime}}$-point in the unit cell of the $b$-von Neumann lattice
determines uniquely the site $\left( \frac{2\pi}{a}t,as\right) $ in the
$a$-von Neumann lattice. This leads to a one-to-one correspondence between the
$M$ $(K_{n^{\prime}},Q_{m^{\prime}})$-points in the unit cell of the $b$-von
Neumann lattice [see Eq.\ (9)] and the $M$ sites $\left( \frac{2\pi}%
{a}t,as\right) $ (with $t=0,1,\dots,M_{1}-1; \ s=0,1,\dots,M_{2}-1$) of the
$a$-von Neumann lattice. A similar analysis shows the one-to-one
correspondence between the $k_{n},q_{m}$-points in the unit cell of the
$a$-von Neumann lattice and the sites of the $b$-von Neumann lattice. From
this we deduce the result stated in the abstract, namely, that the
$kq$-coordinates (we drop the subscripts) in a unit cell of an $a$-von Neumann
lattice are conjugate to the sites of this same lattice. This we deduce from
the fact that the $kq$ and $KQ$ are conjugate coordinates [10], and, as we
have shown, the $KQ$-coordinates determine uniquely the sites of the $a$-von
Neumann lattice. In Fig. 1 the $kq$-coordinates in the unit cell are denoted
by open circles, while the sites of the $a$-von Neumann lattice are denoted by
$\times$'s. Visually, the conjugacy of the $kq$ coordinates and the von
Neumann lattice can be expressed by saying that the open circles and the
$\times$'s are conjugate. A similar statement holds for Fig. 2.

Let us now restate the results in the language of operators and their
eigenfunctions. In the finite phase plane, the eigenfunctions of the
$b$-operators [Eq.\ (8)] are [5,10]
\begin{equation}
\langle x |K_{n^{\prime}},Q_{m^{\prime}}\rangle= \frac{1}{\sqrt{M_{1}}}
\sum^{M_{1}-1}_{\ell=0}\exp(i K_{n^{\prime}}\ell b)\Delta(x-Q_{m^{\prime}%
}-\ell b),
\end{equation}
where $\Delta(x)$ is 1 when $x$ is a multiple of $Mc$ and zero otherwise.
These functions satisfy the eigenvalue equations
\begin{eqnarray}
T(b)\langle x | K_{n^{\prime}},Q_{m^{\prime}}\rangle= e^{i K_{n^{\prime}}%
b}\langle x | K_{n^{\prime}},Q_{m^{\prime}}\rangle\ , \\
\tau(\frac{2\pi}{b})
\langle x | K_{n^{\prime}},Q_{m^{\prime}}\rangle= e^{i Q_{m^{\prime}}%
\frac{2\pi}{b}} \langle x | K_{n^{\prime}},Q_{m^{\prime}}\rangle\
,\nonumber
\end{eqnarray}
where $e^{i K_{n^{\prime}}b}$ and $e^{i Q_{m^{\prime}}\frac{2\pi}{b}}$ are the
eigenvalues of the operators in Eq.\ (8) [see Eq.\ (9)]. These eigenvalues
can, respectively, be replaced by
\begin{equation}
e^{i K_{n^{\prime}}b}=e^{i t \frac{2\pi}{a}b} \ ; \ \ e^{i Q_{m^{\prime}}%
\frac{2\pi}{b}} = e^{isa\frac{2\pi}{b}}
\end{equation}
This follows from Eq.\ (13) because the term $\frac{2\pi}{b}v^{\prime}$ in the
first equation and the term $u^{\prime}b$ in the second equation do not
contribute to Eq.\ (18). Eqs.\ (17) and (18) show that the operators $T(b)$
and $\tau(\frac{2\pi}{b})$ [Eq.\ (8)] have as their eigenvalues not only the
coordinates $K_{n^{\prime}}$ and $Q_{m^{\prime}}$ but also the sites
$\alpha_{st}$ in Eq.\ (12) of the $a$-von Neumann lattice. A similar analysis
can be carried out for the eigenfunctions $\langle x |k_{n},q_{m}\rangle$ of
the operators $T(a)$ and $\tau(\frac{2\pi}{a})$ in Eq.\ (5) [see Eqs.\ (6) and
(7)]. As in Eq.\ (13), we can connect $k_{n}$ and $q_{m}$ in Eq.\ (7) to the
sites of the $a$-von Neumann lattice [see Eq.\ (7)]
\begin{equation}
k_{n} + t\frac{2\pi}{a}=v^{\prime}\frac{2\pi}{b}\\
q_{m} + sa = u^{\prime}b \ ,\nonumber
\end{equation}
where again $s,t,u^{\prime}$ and $v^{\prime}$ are integers, with $s,t$
relating to the $a$-von Neumann lattice, and $u^{\prime},v^{\prime}$ to the
$b$-von Neumann lattice. As before in Eq.\ (13), $k_{n}$ and $q_{m}$ determine
uniquely these integers. We can therefore write equations, similar to
Eqs.\ (18),
\begin{equation}
e^{ik_{n}a}=e^{iv^{\prime}\frac{2\pi}{b}a} \ ; \ \ e^{iq_{m}\frac{2\pi}{a}%
}=e^{iu^{\prime}b\frac{2\pi}{a}}
\end{equation}
This follows from Eq.~(19). We obtain the result that not only $k_{n}$ and
$q_{n}$ but also the sites
\begin{equation}
\beta_{u^{\prime}v^{\prime}}=u^{\prime}b+v^{\prime}\hbar\frac{2\pi}{b}
\end{equation}
of the $b$-von Neumann lattice are eigenvalues of the operators in Eq.~(5).
The results in Eqs.~(18) and (20) are surprising and clearly hold only when
$M_{1}$ and $M_{2}$ in Eq.~(4) are relatively prime.

We are now in a position to write down a new quantum-mechanical wave function,
which has as its variables the sites of a von Neumann lattice. Consider the
wave function $C^{(a)}(k_{n},q_{m})$ [10,11] in the $kq$ representation. By
using
Eq.~(19), that connects $k_{n}$ and $q_{m}$ in Eqs.~(6) and (7) to the sites
of the $a$- and $b$-von Neumann lattices, the wave function $C^{(a)}%
(k_{n},q_{m})$ will become
\begin{equation}
C^{(a)}(k_{n},q_{m})=\exp\left( 2\pi i s v^{\prime}\frac{M_{1}}{M_{2}}\right)
C^{(a)}\left( v^{\prime}\frac{2\pi}{b},u^{\prime}b\right)
\end{equation}
where we have used the periodic boundary conditions satisfied by a
$kq$-function [11]. Eq.\ (22) shows that the $kq$-function, $C^{(a)}%
(k_{n},q_{m})$, on the $a$-von Neumann lattice, determines a wave function
$C^{(a)}\left( v^{\prime}\frac{2\pi}{b},u^{\prime}b\right) $ which has as its
arguments the sites of the $b$-von Neumann lattice. This is a hitherto unknown
function in quantum mechanics. The square of its absolute value
\begin{equation}
|C^{(a)}(v^{\prime}\frac{2\pi}{b},u^{\prime}b)|^{2} %
\end{equation}
gives the probability for a simultaneous measurement of the
momentum $v^{\prime}\hbar\frac{2\pi}{b}$ and the coordinate
$u^{\prime}b$ on the $b$-von Neumann lattice. Similarly, we can
use the $C^{(b)}(K_{n^{\prime}},Q_{m^{\prime}})$ for defining the
wave function $C^{(b)}\left( t\frac{2\pi}{a},sa\right) $. The
square of its absolute value
\begin{equation}
|C^{(b)}(t \frac{2\pi}{a},sa)|^{2}
\end{equation}
gives the probability for a simultaneous measurement of the
momentum $t\hbar\frac{2\pi}{a}$ and the coordinate $sa$ on the
$a$-von Neumann lattice.

In summary, it has been shown that on a finite phase plane of length $Mc$ in
the $x$-direction and $\hbar\frac{2\pi}{c}$ in the $p$-direction, the
$kq$-coordinates and the sites of the corresponding von Neumann lattice form
mutually conjugate coordinates. This result holds when $M$ is a product of two
relatively prime numbers $M_{1}$ and $M_{2}$. By using this conjugacy, a new
wave function was introduced into elementary quantum mechanics, which has as
its arguments the sites of the von Neumann lattice. The square of the absolute
value of this wave function gives the probability of simultaneously measuring
the momentum and coordinate on a von Neumann lattice. This result brings us
back to the work of von Neumann in the very beginning of quantum mechanics
[12], where he was looking for a complete and orthogonal set of eigenfunctions
on a von Neumann lattice. Much work has been done and published on this
subject [13-15], attempting to achieve the von Neumann hypothesis [15] in one
or another approximation or modification. We refer, in particular, to a
recently published paper on the subject with an original approach to the
problem, and which gives also a good summary of the literature [16]. It is
worth noting that the results of this work and of Ref. [10] establish
properties of mutually unbiased bases (MUB) in a space of dimension $d$ which
is not a power of a prime. This may be of interest in the attempts to obtain a
set of $d+1$ MUB for any dimension.


\begin{thebibliography}{99}
%


\bibitem {Schwinger}Schwinger J 1960 \emph{Proc. Nat. Acad. Sci.} USA \textbf{46} 570.

\bibitem {Wootters}Wootters W K 2006 \emph{Found. Phys.} \textbf{36} 112.

\bibitem {Barlett}Bartlett S D, de Guise H and Sanders B C 2002 \emph{\PR A} \textbf{65} 052316.

\bibitem {Leboeuf}Leboeuf P, Kurchan J, Feingold M and Arovas D P 1992 \emph{Chaos} \textbf{2} 125.

\bibitem {Zak}Zak J 1989 \emph{\JMP} \textbf{30} 1591.

\bibitem {Paz}Paz J P 2002 \emph{\PR A} \textbf{65} 062311.

\bibitem {Vourdas}Vourdas A 2004 \emph{Rep. Prog. Phys.} \textbf{67} 267.

\bibitem {Ivanovic}Ivanovi\'{c} I D 1981 \emph{\JPA} \textbf{14}
3241.

\bibitem {Opatrny}Opatrny T, Welsch D G and Buzek V 1996 \emph{\PR A} \textbf{53} 3822.

\bibitem {Mann}Mann A, Revzen M and Zak J 2005 \emph{\JPA} \textbf{38} L389.

\bibitem {Zak1}Zak J 1967 \emph{\PRL} \textbf{19} 1385; Zak J 1970 \emph{Physics
Today} \textbf{23} 51.

\bibitem {Neumann}von Neumann J 1955 \textit{Mathematical Foundations of Quantum
Mechanics} (Princeton Univ. Press, Princeton) pp 406, 407.

\bibitem {Wilson}Wilson K G 1987 \emph{Generalized Wannier functions}, Cornell
University preprint; Sullivan D J, Rehr J J, Wilkins J W and Wilson
K G 1989 \emph{Phase space Wannier functions in electronic structure
calculations} Cornell University Press.

\bibitem {Daubechies}Daubechies I, Jaffard S and Journes J L 1991 \emph{SIAM J.
Math. Anal.} \textbf{22} 559.

\bibitem {Zak2}Zak J 2003 \emph{\JPA} \textbf{36} L553.

\bibitem {Halliwell}Halliwell J J 2005 \emph{\PR A} \textbf{72} 042109.
\end{thebibliography}
\end{document}